\documentclass[aps,prb,twocolumn,superscriptaddress,showpacs,longbibliography]{revtex4-1}  
\usepackage{graphicx}
\usepackage{bm}
\usepackage{color}
\usepackage{amsmath}
\usepackage{tabularx}
\usepackage{ifsym}
\usepackage{amssymb}

\newcommand{\degC}{$^{\circ}$C}

\newcommand{\FeCoSi}{Fe$_{1-x}$Co$_{x}$Si~}

\begin{document}

\title{Manipulation of the spin helix in FeGe thin films and FeGe/Fe multilayers}

\author{Nicholas~A.~Porter}
\affiliation{School of Physics \&\ Astronomy, University of Leeds, Leeds LS2 9JT, United Kingdom}

\author{Charles~S.~Spencer}
\affiliation{School of Physics \&\ Astronomy, University of Leeds, Leeds LS2 9JT, United Kingdom}

\author{Rowan~C.~Temple}
\affiliation{School of Physics \&\ Astronomy, University of Leeds, Leeds LS2 9JT, United Kingdom}

\author{Christian~J.~Kinane}
\affiliation{ISIS, STFC Rutherford Appleton Laboratory, Chilton, Didcot, Oxon. OX11 0QX, United Kingdom}

\author{Timothy~R.~Charlton}
\affiliation{ISIS, STFC Rutherford Appleton Laboratory, Chilton, Didcot, Oxon. OX11 0QX, United Kingdom}

\author{Sean~Langridge}
\affiliation{ISIS, STFC Rutherford Appleton Laboratory, Chilton, Didcot, Oxon. OX11 0QX, United Kingdom}

\author{Christopher~H.~Marrows}\email[email:~]{c.h.marrows@leeds.ac.uk}
\affiliation{School of Physics \&\ Astronomy, University of Leeds, Leeds LS2 9JT, United Kingdom}


\begin{abstract}
  Magnetic materials without structural inversion symmetry can display the Dzyaloshinskii-Moriya interaction, which manifests itself as chiral magnetic ground states. These chiral states can interact in complex ways with applied fields and boundary conditions provided by finite sample sizes that are of the order of the lengthscale of the chiral states. Here we study epitaxial thin films of FeGe with a thickness close to the helix pitch of the helimagnetic ground state, which is about 70~nm, by conventional magnetometry and polarized neutron reflectometry. We show that the helix in an FeGe film reverses under the application of a field by deforming into a helicoidal form, with twists in the helicoid being forced out of the film surfaces on the way to saturation. An additional boundary condition was imposed by exchange coupling a ferromagnetic Fe layer to one of the interfaces of an FeGe layer. This forces the FeGe spins at the interface to point in the same direction as the Fe, preventing node expulsion and giving a handle by which the reversal of the helical magnet may be controlled.
\end{abstract}

\date{\today}
\maketitle

\section{Introduction}

Although non-collinear spin textures have been known for decades\cite{Strandburg1962,Bak1980,Ericsson1981}, the scientific community has been largely preoccupied until recently with the study of collinear ferromagnetic materials. Nevertheless, non-uniform spin textures can arise due to the Dzyaloshinskii-Moriya interaction (DMI) which introduces a chirality to the magnetism and competes with the ferromagnetic (FM) exchange interaction to determine the degree of canting of neighbouring spins and thus the lengthscale of the chiral structures. A non-zero DMI requires a breaking of structural inversion symmetry. This can be achieved artificially in extremely thin FM layers \cite{Heinze2011,Romming2013,Chen2013,Hrabec2014} adjacent to heavy elements where there is broken inversion symmetry at the interfaces. On the other hand, it is also possible in bulk where the unit cell of the crystal lacks inversion symmetry. The B20 structure satisfies the latter criterion, and all of the magnetic materials with this crystal structure exhibit a helimagnetic ground state\cite{Ishikawa1976,Lebech1989,Grigoriev2006,Grigoriev2007,Uchida2006} which may be converted into a skyrmion spin texture upon application of a sufficiently large magnetic field.\cite{Muhlbauer2009,Yu2010,Yu2011}

Of the B20 monosilicides\cite{Manyala2000,Muhlbauer2009,Sinha2014} and monogermanides\cite{Kanazawa2011,Wilhelm2011}, FeGe has the highest magnetic ordering temperature, $T_\mathrm{N} \sim 276$~K,\cite{Wilhelm2011,Ericsson1981,Lebech1989} which is maintained in thin film form,\cite{Huang2012,Porter2014} making it the best available candidate for any future spintronic technologies based on bulk DM interactions. Substantial enhancements of the magnetic ordering temperature have been demonstrated in \FeCoSi\ through the use of epitaxial strain\cite{Sinha2014} and the same opportunity may be present in FeGe to eventually yield room temperature helimagnetism. To integrate B20 materials into existing technologies the material must be available in a thin film form that is amenable to the conventional planar processing methods used in microelectronics manufacturing.  We have grown high quality epitaxial FeGe by molecular beam epitaxy (MBE) using methods similar to that for MnSi\cite{Karhu2011} and \FeCoSi\cite{Porter2012,Sinha2014}. Prior to this work FeGe has previously been grown by high temperature sputtering,\cite{Huang2012,Porter2014} revealing the topological Hall effect arising from the spin textures in the films.

In cubic B20 helimagnets at low fields, the orientation of the helix is determined by the relatively weak cubic anisotropy energy term and, in the case of FeGe, it is oriented along the $\langle 111 \rangle$ cubic axes at zero field. This anisotropy can be overcome by a small magnetic field which aligns the propagation vector of the helix, $\bf{Q}$, to itself. The uniaxial anisotropy, $K_\mathrm{u}$, introduced by shape, the strain in epilayers\cite{Karhu2012,Porter2012,Huang2012,Sinha2014}
and from the surfaces in thinned crystals\cite{Yu2011} is often much greater than the cubic anisotropy that, in these cases, can be neglected. Thus, the direction of $\bf{Q}$ is determined by the uniaxial anisotropy\cite{Karhu2012,Wilson2013}. If the film possesses an easy plane (i.e. a hard axis out-of-plane, such as is provided by shape anisotropy) then the helix propagates normal to the plane and, if $K_\mathrm{u}$ is sufficiently large, reorientation does not occur before saturation. In this case, the helix distorts into a helicoid until eventually the film reaches a fully in-plane magnetized state. Studies of thin epilayers have so far been confined to single layers of helimagnetic material only,\cite{Karhu2011,Karhu2012, Wilson2013} but with thin film growth there is the potential to grow multilayers to create metamaterials with favorable properties compared to isolated films. Here we report on the growth of bilayers of B20 chiral FeGe and ferromagnetic (FM) Fe, and use polarized neutron reflectometry (PNR) to determine how the field-induced reorientation of the spin helix is altered when it is strongly coupled to a FM layer. We show that an FeGe layer reverses its magnetization under field by the distortion of the helix into a helicoid and the expulsion of nodes in the magnetization profile through the film surfaces. On the other hand, adding a FM Fe layer prevents this node ejection mechanism and provides a handle by which the magnetism in the helical FeGe layer may be controlled with a field.

\section{Growth \& Structural Characterization}

\begin{figure}
  \begin{center}
  \includegraphics[width=8cm]{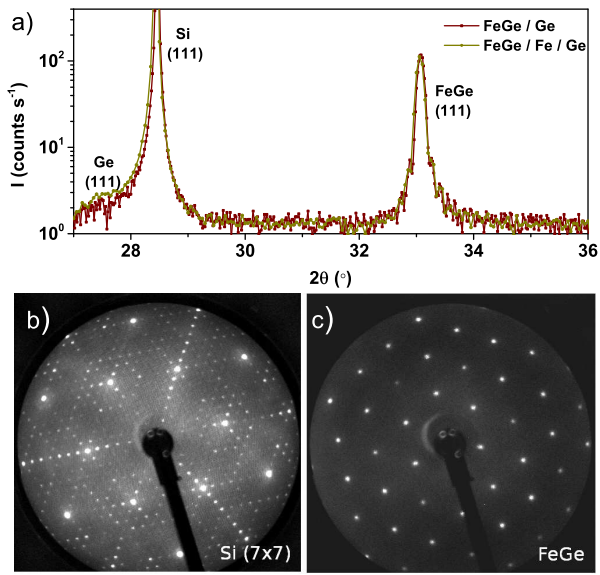}
  \end{center}
  \caption{Crystallographic characterization of the samples. (a) XRD spectra from the the two heterostructures. (b) LEED pattern of annealed Si (111) substrate showing 7 $\times$ 7 surface reconstruction. (c) LEED pattern from the FeGe/Ge film before Ge deposition showing epitaxial growth in the (111) direction with a 30$^\circ$ in -plane rotation with respect to the substrate. \label{fig:xtal}}
\end{figure}

Epilayers of FeGe were grown using molecular beam epitaxy (MBE) by co-deposition from two electron beam sources. 20~mm~$\times$~20~mm pieces of Si (111) wafer were heated to 1200\,\degC\ before cooling to room temperature, whereupon a $7 \times 7$ reconstruction was observed in the low high energy electron diffraction (LEED) pattern, as shown in Fig. \ref{fig:xtal}(b), indicating a clean and ordered Si surface. LEED was then subsequently used to confirm the presence of the B20 phase after the growth at 230\,\degC of nominally 64~nm of FeGe (Fig. \ref{fig:xtal}(c)). This thickness was chosen to be close to the previously measured helix pitch in FeGe.\cite{Lebech1989} The Si-FeGe lattice matching results in the FeGe $[111]$ normal to the surface, as for the Si substrate, but with a 30$^\circ$ in-plane rotation that yields the $[11\bar{2}]$ direction in the film parallel to the Si $[1\bar{1}0]$.\cite{Porter2012,Sinha2014} Auger electron spectroscopy confirmed the equiatomic composition of the FeGe layers in both layers.

The films were then capped at room temperature with Ge (5~nm), or an Fe (4~nm)/Ge (6~nm) bilayer (thicknesses are nominal). The films displayed single Bragg peaks in high angle x-ray diffraction patterns, taken using Cu K$_\alpha$  radiation, as shown in Fig. \ref{fig:xtal}(a), indicating the presence of only the B20 phase with a (111) orientation. A small shoulder on the Si (111) can be seen in the scans arising from reflections from the Ge. In the FeGe/Ge film Pendell\"osung fringes can be seen either side of the FeGe Bragg peak. These fringes are typical of highly ordered growth of smooth layers. The matching of the XRD scans from the two growths is a strong indication of the reproducibility of the film growth by MBE.

\begin{table*}
\caption{X-ray characterisation data: $a_\mathrm{FeGe}$ is the out-of-plane lattice constant determined from the XRD; $t_X$ is the thickness of layer $X$ determined from the fits to the XRR. \label{tab:thickness}}
\begin{ruledtabular}
\begin{tabular}{l|cccc}
Sample         & $a_\mathrm{FeGe}$ (pm) & $t_\mathrm{FeGe}$ (nm) & $t_\mathrm{Fe}$ (nm) & $t_\mathrm{Ge}$ (nm)   \\ \hline
FeGe/Ge      & 469.38 $\pm$ 0.03 & 67.8 $\pm$ 0.1       & -             & 4.77 $\pm$ 0.07 \\
FeGe/Fe/Ge   & 467.58 $\pm$ 0.03 & 64.2 $\pm$ 0.4       & 5.4 $\pm$ 0.2 & 2.1 $\pm$ 0.1
\end{tabular}
\end{ruledtabular}
\end{table*}

\begin{figure}
  \begin{center}
  \includegraphics[width=8cm]{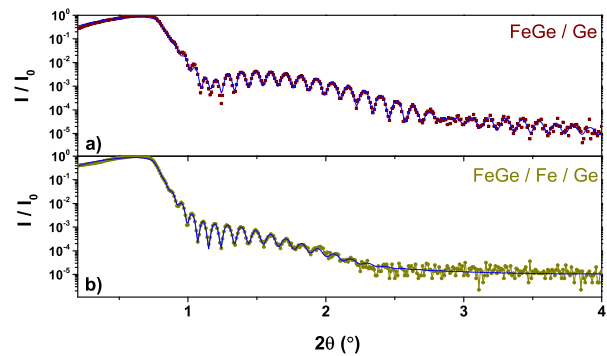}
  \end{center}
  \caption{XRR from the two heterostructures. (a) XRR from FeGe/Ge bilayer. (b) XRR from the FeGe/Fe/Ge trilayer. The fit to the XRR data is shown by the solid line. \label{fig:xrr}}
\end{figure}

X-ray reflectometry (XRR) was used to determine the film thicknesses. The XRR data are shown in Fig. \ref{fig:xrr}. The presence of well-defined Kiessig fringes indicates that the layers are flat and sharply defined. The data were fitted with the GenX code\cite{Bjorck2007} to yield structural parameters describing the layer stack of each sample. The fitted layer thicknesses are shown in Table \ref{tab:thickness} and are all close to the nominal values. All fitted interface widths (quadrature sum of roughness and intermixing) were less than 1~nm with the exception of the Si wafer surface ($1.70 \pm 0.09$~nm) and top surface of the Ge ($1.6 \pm 0.9$~nm) in the sample containing the Fe layer. The fitted layer densities were all within 10~\%\ of the bulk values.

\section{Magnetometry}

Magnetometry was performed in a SQUID-vibrating sample magnetometer with the magnetic field applied parallel to the FeGe $[110]$ direction in the film plane. We define a dynamic susceptibility, $\chi = \mu_0^{-1} dm/dH$, derived from numerical differentiation of the $m(H)$ curves, where $m$ is the film moment and $H$ is the applied magnetic field.

\begin{figure}[tb]
  \includegraphics[width=8.5cm]{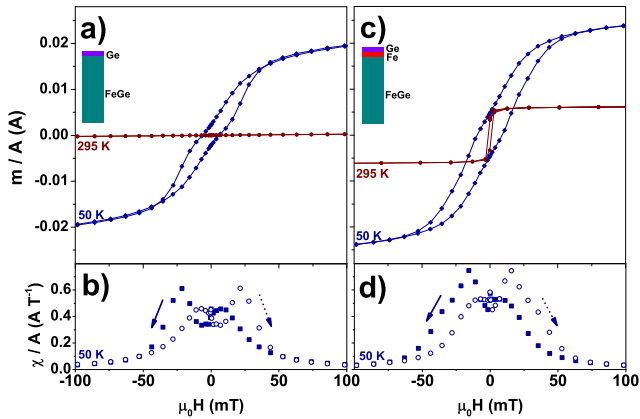}
  \caption{Hysteresis of the magnetic moment $m$ per area $A$ for the two multilayers. (a) Hysteresis of FeGe/Ge above and below $T_\mathrm{N}$. The FeGe film is paramagnetic at 295~K, but shows hysteresis at 50~K.  (b) Dynamic susceptibility $\chi$ per area derived from the 50~K data in (a) with open and closed symbols corresponding to the two field sweep directions. (c) Hysteresis of FeGe/Fe/Ge is observed at 295~K from the Fe layer, leading to a slight modification of the 50~K data with respect to the FeGe/Ge sample. The 50~K hysteresis loop is numerically differentiated in (d).\label{fig:vsm}}
\end{figure}

Magnetic hysteresis loops for each of the two structures, measured above and below $T_\mathrm{N}$, are shown in Fig. \ref{fig:vsm}. In Fig. \ref{fig:vsm}(a), at $T = 295$~K ($T > T_\mathrm{N}$), the FeGe in the FeGe/Ge bilayer was paramagnetic with a field-independent susceptibility of $\chi = 0.038~\mu_\mathrm{B}$~atom$^{-1}$ T$^{-1}$. In contrast, below $T_\mathrm{N}$ at 50~K, the FeGe in the bilayer was magnetically ordered and a hysteretic magnetization loop was measured. This hysteresis suggests an irreversible unwinding of the helix through to the field-polarized state. The saturation magnetization for $\mu_0 H \gtrsim 1$~T was 360~kA/m corresponding to $\sim0.5~\mu_\mathrm{B}$ atom$^1$, roughly half the value of 1~$\mu_\mathrm{B}$ per Fe atom quoted for bulk FeGe.\cite{Yamada2003,Pedrazzini2007} The susceptibility per unit area of the multilayer was obtained from the derivative of the data in Fig. \ref{fig:vsm}(a), and is plotted in Fig. \ref{fig:vsm}(b). Two peaks are observed in each branch of the $\chi$ hysteresis loop, one occurring before zero field and one afterwards. The presence of two peaks suggests that the distortion of the helix as a function of magnetic field occurs in a two step process.\cite{Wilson2013}

The hysteresis of the FeGe/Fe/Ge multilayer is shown in Figs. \ref{fig:vsm}(c) and \ref{fig:vsm}(d). At 295~K the FeGe is above its ordering temperature whereas the Fe layer is not and a narrow square hysteresis loop was measured, corresponding to the easy axis switching of Fe layer alone. At 50~K the FeGe layer is also magnetically ordered and clean switching of the Fe layer is not seen in the overall hysteresis. This lack of switching at small fields suggests strong coupling with the interface magnetism in the FeGe layer. Unlike the FeGe/Ge bilayer, this film exhibits, within the resolution of the experiment, a single peak in $\chi$ (Fig. \ref{fig:vsm}(d)) at $\sim 16$~mT. The second peak is reduced to a step-like shoulder on the first. This coupling between the layers can be expected to alter the response to a field in the switching of the helix, which we now explore using PNR.


\section{Polarized Neutron Reflectometry}

We used time-of-flight PNR on the PolRef instrument at ISIS to determine the magnetic depth profile of the samples as the field was varied.\cite{Blundell1992} The FeGe epilayers were subjected to a magnetic field applied parallel to the FeGe $[110]$ direction in the film plane, with the neutron scattering plane orthogonal to the field. The intensity of scattered neutrons of each spin, $I_{+(-)}$ was measured as a function of scattering vector, $q_\mathrm{z} = (4\pi / \lambda) \sin\theta$ where $\theta$ is the incident angle and $\lambda$ is the wavelength of the incident neutrons. The range of $q_\mathrm{z}$ was provided by the distribution of neutron velocities (and hence wavelengths) in the time-of-flight geometry used at PolRef. As the neutron scattering potential is the sum of both nuclear and magnetic scattering,\cite{Zabel2003} by simultaneously fitting both $I_+$ and $I_-$ it is possible to determine the nuclear, $\rho_\mathrm{s}$, and the magnetic, $\rho_\mathrm{m}$, scattering length density (SLD) profiles. Once again, the GenX software\cite{Bjorck2007} was used to fit the PNR data, with the structural parameters used in the neutron fit constrained to be consistent with those determined by XRR. No layer had to have its thickness adjusted from the XRR value by more than 2~nm, and the FeGe thicknesses were the same to within as low a tolerance as 2~\%. The magnetic scattering length density (MSLD) depth profile was then calculated from the magnetic SLD as $M = \rho_\mathrm{m} / (p n)$ where $n$ is the atomic density and $p = 2.95$~fm/$\mu_\mathrm{B}$.

\begin{figure}[tb]
  \includegraphics[width=8.0cm]{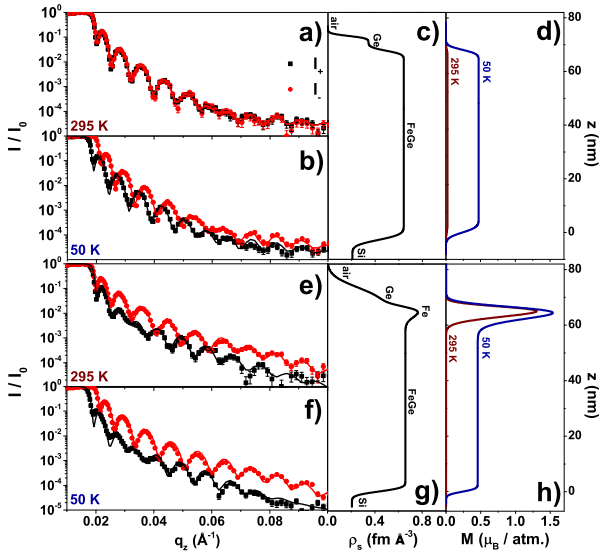}
  \caption{PNR spectra and fitted depth profiles from the two samples FeGe/Ge (a-d) and FeGe/Fe/Ge (e-h) shown above (295~K) and below (50~K) $T_\mathrm{N}$ in a 667~mT field. PNR for the up, $I_{+}$ (squares), and down, $I_{-}$ (circles), spin-polarized neutrons at 295~K shows small splitting arising from paramagnetism only (a) and a large splitting in the saturated magnetic state at 50~K (b). The structural SLD ascertained from the fits (lines in (a) and (b)) is shown in (c) and the magnetic profile at each temperature is shown in (d). PNR for FeGe/Fe/Ge is shown in (e) and (f) with the structural SLD in (g) and magnetic profiles in (h).  \label{fig:satpnr}}
\end{figure}

PNR spectra at the highest available field of 667~mT are shown in Fig. \ref{fig:satpnr} for both samples at equivalent temperatures to the magnetometry. As shown by the $m(H)$ loops in Fig. \ref{fig:vsm}, this is large enough to fully saturate both samples. For the FeGe/Ge bilayer, the reflected intensities from `up' and `down' polarized neutrons are shown in Fig. \ref{fig:satpnr}(a) and Fig. \ref{fig:satpnr}(b) for 295~K and 50~K respectively. The structural SLD as determined from fitting these data (solid lines) is shown in Fig. \ref{fig:satpnr}(c). At 295~K there is only a small separation in the two reflectivity spectra in Fig. \ref{fig:satpnr}(a) suggesting a tiny average moment on the FeGe atoms corresponding to the weak alignment of moments by field in the paramagnetic phase of the film. At the measurement field of 667~mT, the fitting required a uniformly magnetized FeGe film to account for the MSLD with a small average moment of $0.025~\mu_\mathrm{B}~\mathrm{atom}^{-1}$, consistent with the paramagnetic moment measured by SQUID-VSM. At 50~K (and 667~mT) the film is expected to be uniformly magnetized ($M/M_\mathrm{s}\sim0.995$) and, as seen in in Fig. \ref{fig:satpnr}(b), there is a clear separation in the reflectivity spectra. The MSLD used to fit this data is a uniformly magnetized film with $0.47~\mu_\mathrm{B}~\mathrm{atom}^{-1}$, once again in agreement with the magnetometry.

For the FeGe/Fe/Ge trilayer, the PNR in Fig. \ref{fig:satpnr}(e-h) tells a similar story. At 295~K there is significant splitting in the PNR (Fig. \ref{fig:satpnr}(e)) but this is all attributed to the large moment on the Fe atoms in the middle layer (see fitted MSLD in Fig. \ref{fig:satpnr}(g)). At 50~K the PNR differs from that at 295~K, with a clear separation of the two reflections at the critical edge of total external reflection ($q_\mathrm{z} = 0.02$~\AA$^{-1}$). In Fig. \ref{fig:satpnr}(h), the fitted MSLD shows that at 50~K, in addition to there being a slightly higher moment on Fe, the FeGe film is saturated with a magnetization of $0.46~\mu_\mathrm{B}~\mathrm{atom}^{-1}$. Thus, we can see that the FeGe in two multilayers behaves very similarly under temperature variations. Moreover, with temperature we can `turn off' the helimagnetism in the FeGe allowing us to obtain the structural information from the films. The structural profile and saturation magnetization values derived from these fits were used in subsequent fitting at lower fields, where only moment directions were changed to simulate the data acquired as the samples were taken around their hysteresis loop.

\begin{figure}[tb]
  \includegraphics[width=8.0cm]{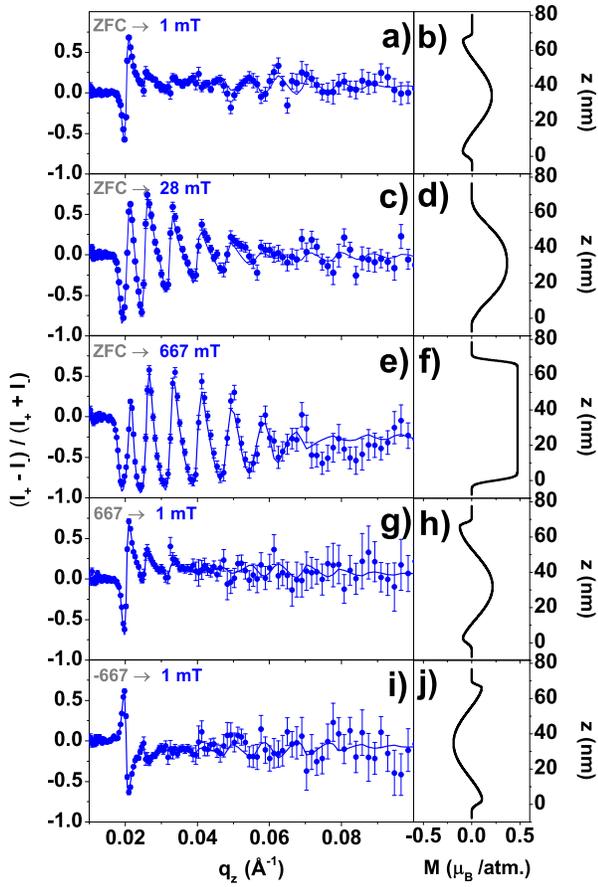}
  \caption{Snapshots of the helicoid reversal in the FeGe/Ge bilayer, measured at 50~K. The SA (a,c,e,g,i) and extracted magnetic profiles (b,d,f,h,j) are displayed for various measurement fields and field histories. After zero field cooling, the SA in a 1~mT field is modeled by a slightly distorted helix with an antinode in the centre of the film (a,b). At 28~mT (c,d) the helix is distorted further into field direction. Saturation occurs at 667~mT (e,f). After positive saturation, at 1~mT (g,h) the moments relax to the initial state. After negative saturation, the moments relax at 1~mT into an inverted state. The solid line in each SA plot is the fit to the PNR spectrum generated by the accompanying magnetic profile. \label{fig:fegehyst}}
\end{figure}

First, we address the behaviour of the FeGe/Ge sample as the field was swept. PNR snapshots taken at various points on the hysteresis loop are shown in Fig. \ref{fig:fegehyst}. For clarity and brevity, the spin asymmetry (SA), defined as $(I_{+} - I_{-}) / (I_{+} + I_{-})$, derived from the PNR spectra are shown rather than the spectra themselves. The film was zero field cooled (ZFC) from room temperature through $T_\mathrm{N}$ to 50~K and then a small 1~mT field was applied for the first measurement. This field was the minimum requirement to maintain the quantization axis of the polarized neutrons. In order model the magnetic profile and to fit the MSLD a distorted helicoid model was used:\cite{Rossler2011,Wilsonthesis2013}
\begin{equation}\label{eqn:Helix1}
    M(z) = M_0 + M_1\sin\left( \frac{2\pi z}{\Lambda} + \phi_0 \right) + M_2\cos^2\left( \frac{2\pi z}{\Lambda} + \phi_0 \right)
\end{equation}
where $\Lambda$ is the wavelength of the undistorted ground state helix, $\phi_0$ determines the translation of the nodes along the $z$ axis and $M_{0}$ gives the offset of the magnetization. For instance, the fits used in Fig. \ref{fig:satpnr} correspond to fitting for $M_1=M_2=0$ to give a saturated state, whilst for the undistorted helix that is expected to be the ground state, $M_0=M_2=0$.

The SA was fitted using a profile with the model from Eq. \ref{eqn:Helix1}. It is important to note that this fitted magnetization profile is actually a spin density wave that only takes account of the magnetization component along the field direction. The transverse component is assumed to be zero. This is because thin films such as ours are racemic and contain grains of both B20 crystal chiral handednesses,\cite{Karhu2010,PorterArXiv2013} which span the height of the film and are typically a few hundreds of nm across. These regions will have helical ground states of opposite handedness, since the sign of the DMI is defined by the crystal chirality in the B20 materials. Since the in-plane coherence length of the neutron beam is typically several $\mu$m, large numbers of grains are sampled coherently. Provided that the helices in the left and right handed grains have equal values of $\phi_0$, the transverse components of magnetization will cancel and only the longitudinal components will remain, yielding the spin density wave expressed by Eq. \ref{eqn:Helix1}. This spin density wave can be described in terms of nodes and antinodes, which is a language that does not apply strictly to the helices, but is nevertheless convenient to use in order to discuss the magnetic depth profiles that the PNR reveals. A node in the spin density wave represents the point in $z$ where the spins in the helix are orthogonal to the field.

\begin{figure}[tb]
  \includegraphics[width=8.0cm]{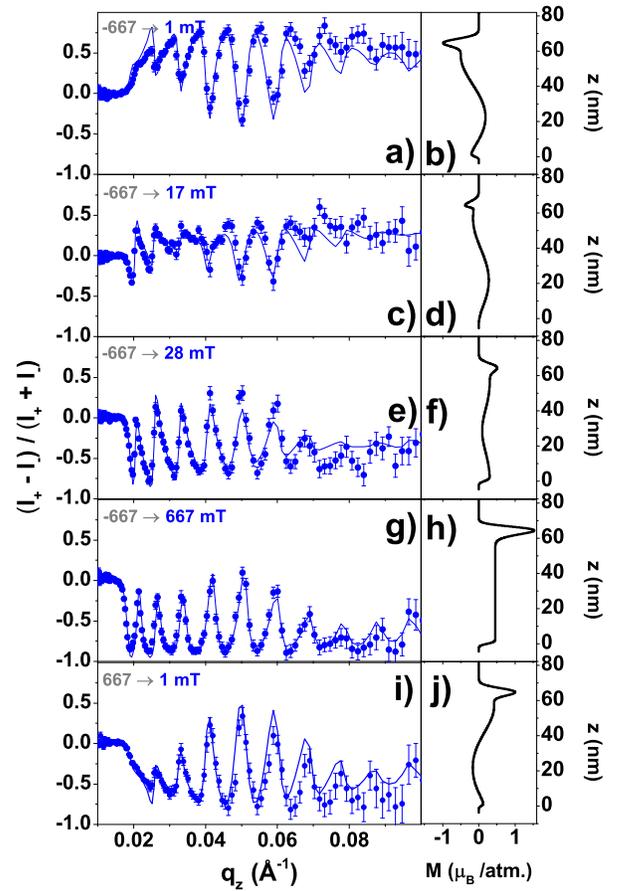}
  \caption{Snapshots of the helicoid reversal in the FeGe/Fe/Ge trilayer, measured at 50~K. The SA (a,c,e,g,i) and extracted magnetic profiles (b,d,f,h,j) are displayed for various measurement fields and field histories. After negative saturation, in a 1 mT field the Fe cap is aligned anti-parallel to the applied field and the antinode of the FeGe layer is shifted towards the bottom of the film (a,b). In 17 mT the Fe layer begins to switch and with it the spin helix (c,d). Once the Fe is starting to become field aligned at 28 mT (e,f) the helix distorts further, and  saturated is then reached at 667 mT (g,h). After positive saturation, the profile of moments in film at 1 mT (g,h)are of opposite sign to those in (a,b). The solid line in each SA plot is the fit to the PNR spectrum generated by the accompanying magnetic profile. \label{fig:fegefehyst}}
\end{figure}

The measured SA (points) at 1~mT and fit (solid line) are shown in Fig. \ref{fig:fegehyst}(a) with the profile used for the fit in Fig. \ref{fig:fegehyst}(b). After ZFC and 1~mT field applied there existed a helicoid state with $\Lambda = 70$~nm with an antinode in the center of the layer. The spins in the antinode are aligned in the field direction, and so we can conclude that the small 1~mT field required to polarize the neutrons is enough to break the degeneracy between this magnetic profile and one with an oppositely polarized antinode: see Fig. \ref{fig:fegehyst}(j). We found that the helical pitch in this (relatively weakly) strained film to be $\Lambda = 70 \pm 5$~nm, in excellent accord with that observed in the bulk\cite{Ericsson1981,Lebech1989,Yu2011,Shibata2013}. Increasing the field to 28~mT (Fig. \ref{fig:fegehyst}(c)) began to distort the helix into a helicoid, pulling the spins at the interfaces perpendicular to the applied field as shown in \ref{fig:fegehyst}(d). At 667~mT (Fig. \ref{fig:fegehyst}f) the magnetic profile was saturated. Reducing the field back to 1~mT (Fig. \ref{fig:fegehyst}(g) and (h)) returned to the original configuration in Fig. \ref{fig:fegehyst}b suggesting that a minor field sweep without crossing the coercive field of the multilayer film is reversible. To demonstrate the irreversible behaviour when the coercivity is crossed, we took the system to negative saturation ($-667$~mT) and then repeated the measurement at 1~mT (Fig. \ref{fig:fegehyst}{i}), with the fit yielding an inverted profile, displayed in Fig. \ref{fig:fegehyst}(j), to the original state that is shown in Fig. \ref{fig:fegehyst}(b).


We then measured the FeGe/Fe/Ge trilayer with a slightly different field history, chosen to reflect its modified hysteresis loop and the fact that the Fe layer is already magnetized at the start of the cooling process. After beginning from a negatively saturated state and then applying a 1~mT field (Fig. \ref{fig:fegefehyst}(a)), the Fe layer was still magnetized in the negative direction, as shown in Fig. \ref{fig:fegefehyst}(b). The coupling between the Fe and the FeGe was strong enough to bias the upper interface of the FeGe, deforming the helical state and pushing the top node deeper into the film than in the case of the free helix shown in Fig. \ref{fig:fegehyst}(b). We then measured at an increased intermediate field of 17~mT, with SA data shown in Fig. \ref{fig:fegefehyst}(c). At this field, in the fitted profile shown in Fig. \ref{fig:fegefehyst}(d), the Fe layer magnetization is still negative but of a reduced value, due to the fact that it was beginning to switch and this value represents the average over a domain structure. This is the origin of the large susceptibility shown in Fig. \ref{fig:vsm}(d) at the peak. This switching process was still continuing at 28~mT (Fig. \ref{fig:fegefehyst}(e) where the Fe layer has now reversed its average direction of magnetization but the overall magnetization remains small (Fig. \ref{fig:fegefehyst}(f)). During the switching process the neighboring FeGe spins have retained their alignment with the Fe layer due to the strong interfacial exchange coupling. This coupling maintains the relative orientation of the distorted helicoid to the Fe layer, and the entire helicoid also switches at this point. At 667~mT (Fig. \ref{fig:fegefehyst}(g), the entire system is saturated, as shown in Fig. \ref{fig:fegefehyst}(h). After this saturation in a positive field, returning to 1~mT (Fig. \ref{fig:fegefehyst}(j)) the film completes a hysteresis loop and the opposite state to the original state in Fig. \ref{fig:fegefehyst}(b) is generated.

\section{Discussion \& Conclusions}

These data show how a magnetic helix responds to the application of a field under different boundary conditions. When the helix is confined to a layer that is just smaller than its wavelength, its equilibrium state was determined by PNR to be that with an antinode in the centre of the layer, separated by nodes from regions of opposing spins near to each interface. The symmetry of the magnetization profile about the center of the film is enforced by the symmetry of the two interfaces. Application of a field twists the state to push the nodes out of the film surfaces, which is the process that corresponds to the larger peak in $\chi$ in the forward field direction. Once the nodes are ejected the magnetization can be saturated by higher fields. The weaker peak in $\chi$ that is observed before zero field is reached can be assigned to the inverse process: the nodes re-entering the film as the system returns from saturation.

When a further magnetic boundary condition is imposed by the addition of the Fe layer, this has two effects: it breaks the symmetry between the two interfaces, which means that the helicoidal state need no longer be symmetric about the center of the film; and it exchange couples to the spins at that interface, which means that they must track the much larger magnetization of the Fe layer as it follows the field. This also prevents a node being expelled through this interface. This leads to a natural explanation for the modification of the field dependence of $\chi$, where the peak feature on applying a forward field corresponds to the reversal of the Fe layer and node expulsion and re-entry is suppressed. Rather, the single peak arises when the Fe Layer switched, dragging the helicoidal state in the FeGe with it. The large Fe moment becomes a convenient handle by which the helical magnetism in the FeGe layer may be manipulated by a field.

\begin{acknowledgments}
The authors thank Dr M. Ali and G. Burnell for useful discussions.  We acknowledge financial support from the UK EPSRC (grant numbers EP/J007110/1, EP/K00512X/1, and EP/J021156/1).
\end{acknowledgments}

\bibliography{FeGe}

\end{document}